\documentclass[proof]{WileyASNA-v1}

\articletype{Article Type}%

\received{}
\revised{}
\accepted{}

\begin{document}

\title{Black hole in asymmetric cosmological bounce}

\author[1]{Daniela P\'erez*}

\author[2]{Santiago E. Perez Bergliaffa}

\author[1,3]{Gustavo E. Romero}

\authormark{P\'erez et al. }

\address[1]{\orgdiv{Instituto Argentino de Radioastronom\'ia}, \orgname{IAR, CONICET/CIC/UNLP}, \orgaddress{\state{Buenos Aires}, \country{Argentina}}}

\address[2]{Departamento de F\'isica Te\'orica, Instituto de F\'isica,
Universidade do Estado de Rio de Janeiro,
CEP 20550-013, Rio de Janeiro, Brazil}

\address[3]{Facultad de Ciencias Astron\'omicas y Geof\'isicas, Universidad Nacional de La Plata, Paseo del Bosque s/n, 1900 La Plata, Buenos Aires, Argentina}


\corres{*Camino Gral. Belgrano Km. 40, C.C.5, (1894) Villa Elisa, Buenos Aires, Argentina \email{danielaperez@iar.unlp.edu.ar}}

\abstract{We determine the causal structure of the McVittie spacetime for a cosmological model with an asymmetric bounce. The analysis includes the computation of trapping horizons, regular, trapped, and anti-trapped regions, and the integration of the trajectories of radial null geodesics before, during, and after the bounce. We find a trapped region since the beginning of the contracting phase up to shortly before the bounce, thus showing the existence of a black hole.  When the universe reaches a certain minimum scale in the contracting phase, the trapping horizons disappear and the central singularity becomes naked. These results suggest that neither a contracting nor an expanding universe can accommodate a black hole at all times.}


\maketitle

\section{Introduction}

In 1933, McVittie \citep{mcv33} discovered an exact solution of Einstein's field equations that describes an inhomogeneity embedded in a Friedmann-Lema\^itre-Robertson-Walker (FLRW) cosmological background. The line element takes the form:
\begin{eqnarray}\label{mcv}
 ds^{2}  & = & - f(t,R) dt^{2} - \frac{2 H(t) R}{\sqrt{1- 2m_0/R}} dt dR + \frac{dR^{2}}{1- 2m_0/R}\nonumber\\
& +&  R^{2}\left( d\theta^{2} +  \sin^{2}{\theta} d\phi^{2}\right),
\end{eqnarray}
where
\begin{equation}
f(t,R) \equiv 1- 2m_0/R - H(t)^2 R^2.   
\end{equation}
Here, $H(t) \equiv \dot{a}(t)/a(t)$ is the Hubble factor corresponding to the
background cosmological model, and $m_0$ is a non-negative constant. The coordinate $R$ denotes the aereal radius coordinate.\footnote{The areal radius coordinate $R$ is defined by $R:= \sqrt{\mathcal{A}/4\pi}$, where $\mathcal{A}$ is the area of the 2-sphere of symmetry, and where
\begin{equation}
 d^{2}\Omega_{(2)} := {d\theta}^{2} + {\sin{\theta}}^{2}  {d\phi}^{2},
\end{equation}
is the line element on the unit 2-sphere \cite{far15}.} Setting $a(t) \equiv 1$,  Eq. \eqref{mcv} reduces to the Schwarzschild line element, and in the limit $m_0 \rightarrow 0$ the FLRW metric is recovered.

The McVittie solution is based on two hypotheses:
\begin{itemize}
\item The matter is a perfect fluid with density $\rho$ and isotropic pressure $p$. The energy-momentum tensor is given by \cite{car+09}:   
\begin{equation}\label{eq: cond1}
\mathbf{T} = \rho  \mathbf{\tilde{u}} \otimes \mathbf{\tilde{u}} + p \left(\mathbf{\tilde{u}} \otimes \mathbf{\tilde{u}} - \mathbf{g}\right),
\end{equation}
where $\mathbf{\tilde{u}}:= \mathbf{g}(\mathbf{u}, .)$ is the 1-form metric dual to the vector $\mathbf{u}$ that represents the four-velocity of the fluid. 

\item The four-velocity of the fluid is
\begin{equation}\label{eq: cond2}
\mathbf{u} = \mathbf{e}_{0},  
\end{equation}
that is, the fluid has zero velocity with respect to the chosen reference frame. Here, $\mathbf{e}_{0}$ is the 0-component of the orthormal tetrad $\lbrace\mathbf{e}_{\mu}\rbrace_{  \mu \in \{0,..,3\}}$ of the metric defined by $\mathbf{e}_{\mu} := {\Vert \mathbf{\partial}/ \mathbf{\partial}x^{\mu} \Vert}^{-1} \mathbf{\partial}/ \mathbf{\partial}x^{\mu}$. 
\end{itemize}
We stress that no equation of state is assumed.
 
After a long debate in the literature about the physical interpretation of the McVittie spacetime, it has been established that the McVittie metric represents a dynamical black hole embedded in a cosmological background (see for instance the series of works by \cite{nol98,nol99a,nol99b,kal+10,lak+11}). The details of the solution and its possible analytical extension depend of
the behaviour of $H(t)$ for $t\rightarrow\infty$ \citep{lak+11}. The solution  
displays a curvature singularity at $R=2m_0$ for finite values of $t$, as evidenced by the Ricci scalar, given by: 
\begin{equation}
{\cal R}=12H^2+\frac{6\dot H}{\sqrt{1-\frac{2m_0}{R}}}.
\end{equation}
The singularity is spacelike and, as shown by \cite{nol99b}, gravitationally weak \citep{tip77}.

The McVittie solution has been investigated, so far, for a standard prescription of the scale factor of the universe. In this work we extend the research to models that allow for an asymmetric bounce. We discuss whether the solution includes a black hole before the bounce and what happens with the horizons along the cosmological history of a bouncing universe. The results we have found, along with those presented by P\'erez et al. \citep{per+20}, will help to obtain a better understanding of both the McVittie solution and the fate of a dynamical black hole through a cosmic bounce.   

The article is organized as follows: in Section \ref{sec:2} we introduce the scalar factor of the asymmetric bouncing cosmological model. In the next section, we compute the trapping horizons and integrate the trajectories of null radial geodesics of the metric; we also provide several plots showing the results. In Subsection \ref{before-bounce}, we analyze the causal structure of the McVittie solution before the bounce, and determine whether a black hole is present. In the last section, we discuss the implications of the results here obtained, and the open issues that deserve to be explore in future works.

\section{Scalar factor for an asymmetric bouncing cosmological model}\label{sec:2}

Cosmological models that display a bounce solve by construction the initial singularity problem, as well as the horizon and flatness problems of the standard cosmological model
\footnote{See  \cite{Novello2008} for a review.}. Models with a bounce
join a contracting phase, in which the universe was very large and almost flat initially,
to a subsequent expanding phase. The bounce can be either generated classically (see e.g.  \cite{Wands:2008tv, Ijjas:2016tpn, Galkina:2019pir}), 
or by quantum effects (see e.g. 
\cite{Peter:2008qz, Almeida:2018xvj, Bacalhau:2017hja, Frion:2018oij}). In this work, we explore the effects of an asymmetric bounce on the McVittie solution. The expression for the corresponding scale factor is given by
\begin{eqnarray}
 a(t) & = & \left[ - 3 p \frac{(1-w)}{4} t + {a_{b}}^{\frac{3 (1-w)}{2} } f(t)  \right]^{\frac{2}{3(1-w)}}, \label{scale-factor}\\
f(t) & = & \left[ 1 + \left(\frac{t}{\sigma^2}\right)^2
 +  \left(3 p \frac{(1-w)}{4}\right)^{2}   \frac{\left(t^{2} + \sigma^{4}\right)}{{a_{b}}^{3 (1-w)} }\right]^{1/2},
\end{eqnarray}
where $w$ is the equation of state parameter of the cosmological fluid, $\sigma$ is the standard deviation of the initial wave function, $p$ is a parameter directly related to the intensity of the asymmetry, and $a_{b}$ is a normalization constant. The expression for the scale factor was obtained by \cite{del+20}  considering quantum corrections to the classical evolution of the scale factor. The corrections were obtained by solving the Wheeler-deWitt equation in the presence of a single perfect fluid, in the framework of the de Broglie-Bohm quantum theory \citep{Pinto-Neto:2013toa}. Asymmetric bouncing models were derived by considering initial wave functions with non null phase velocity.

In what follows, we analyze the causal structure of the McVittie spacetime for two different values of the parameter $p$. In all the calculations, we assume that the background cosmological fluid is radiation dominated, i.e., $w = 1/3$.


\section{McVittie in asymmetric bouncing cosmology}


\subsection{Trapping horizons and null geodesics}

Trapping horizons are defined as the surfaces where null geodesics change their focusing properties \citep{hay94}. Mathematically, this kind of horizon is determined by the condition
\begin{equation}\label{th}
 \theta_{\mathrm{in}} \theta_{\mathrm{out}} = 0, 
\end{equation}
where $\theta_{\mathrm{in}}$ stands for the expansion of ingoing radial null geodesics while $\theta_{\mathrm{out}}$ denotes the expansion of outgoing radial null geodesics, respectively. Regions where $\theta_{\mathrm{in}} \theta_{\mathrm{out}} < 0$ are called \textit{regular}. In the opposite case, $\theta_{\mathrm{in}} \theta_{\mathrm{out}} > 0$, the region is called \textit{anti-trapped} if $\theta_{\mathrm{in}} > 0$ and $\theta_{\mathrm{out}} >0$, and \textit{trapped} if $\theta_{\mathrm{in}} < 0$ and $\theta_{\mathrm{out}} < 0$.

We compute the trapping horizons for the line element \eqref{mcv},  and  Hubble factor given by\footnote{We shall use geometrized units $G = c = 1$. }:
\begin{equation}\label{hubble-factor}
 H(t) =   - \frac{6 \; {a_{b}}^{3 w/2} \; p \; (-1 + w) \left(t^2 + \sigma^4\right) + 2 \; a_{b}^{3/2} \; t \; q(t)}{3 \; (-1 + w) \left(t^2 + \sigma^4\right) \left(3 \; a_{b}^{3 w/2} \; p  \; t  \; (-1+w) +{a_{b}}^{3/2} q(t)  \right) },
\end{equation}
where,
\begin{equation}
q(t) = \sqrt{\left(9 {a_{b}}^{-3+w} p^2 \left(-1+w\right)^2 + \frac{16}{\sigma^4}\right)\left(t^2 + \sigma^4\right)}  
\end{equation}
in the time interval $- \infty < t < + \infty$. The condition \eqref{th} is equivalent to:
\begin{equation}\label{trap-McVittie} 
  \tilde{g}(t,R) = H(t)^{2}R^{3} - R + 2m_0 =0.
  \end{equation}
  
  The properties of the trapping horizons depend on the values of the parameters $m_0$, $p$, $a_b$, $w$ and $\sigma$. Since we attempt to asses the effects of the asymmetric bounce on the McVittie solution, we assume for the sake of simplicity, $m_0 = a_{b} = \sigma = 1$ and consider separately the cases $p = 1$ and $p = 3$. 
  
  \subsubsection{Case $p = 1$}
  
\begin{figure}[htb]
\centering
\includegraphics[width=70mm,height=70mm]{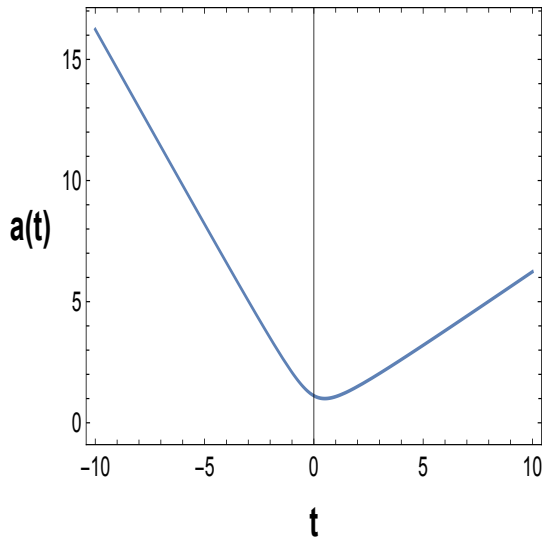} 
\caption{Plot of the scale factor as a function of time for $p = 1$. Here, $m_0 = a_{b} = \sigma = 1$ and $w = 1/3$.} \label{fig1}
\end{figure}

  We show in Figure \ref{fig1} the plot of the scale factor given by \eqref{scale-factor} for $p = 1$. We clearly see that the bounce is asymmetric. The slope in the contracting phase is larger than the slope in the expanding phase. The bounce occurs for the $t_{b}= 0.5 $. 
  
\begin{figure}[htb]
\centering
\includegraphics[width=80mm,height=80mm]{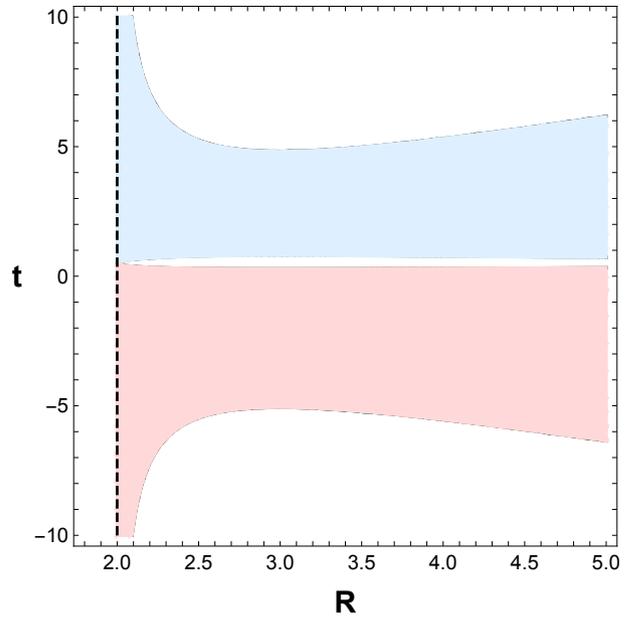} 
\caption{The black lines denotes the trapping horizons. The dot-dashed curve corresponds to the singularity. The white, red light and blue light  zones indicate the regular, trapped, and anti-trapped regions of the spacetime, respectively. Here $p =1$, $m_0 = a_{b} = \sigma = 1$ and $w = 1/3$. } \label{fig2}
\end{figure}

 In Figure \ref{fig2}  we show the trapping horizons and the regular, trapped, and anti-trapped regions of the spacetime. The black lines indicate the location of the trapping horizons. Very close to $t_{b}$, just before and after the bounce, there is a trapping horizon. There are two additional trapping horizons in the contracting and expanding phase, respectively.  In the expanding phase, there is a moment in time $t_{\mathrm{af}} = 4.82$ when a single trapping horizon at $R_{\mathrm{af}} = 3 m_0$ begins to exist and immediately after an inner $R_{-}$ and outer $R_{+}$ trapping horizons emerge. In the contracting region there are also an inner $R_{-}$ and outer $R_{+}$ horizons for large negative values. As time increases,  $R_{-}$ becomes larger and $R_{+}$ becomes smaller up to $t_{\mathrm{bf}} = -5.06 $ when they fuse into one at $R_{\mathrm{bf}} = 3m_0$ . In the limit $t \rightarrow \pm \infty$, $R_{-} \rightarrow 2 m_{0}$. The dot-dashed curve marks the location of the singularity $R = 2 m_0$ for $t$ finite.

The second root of Eq. \eqref{trap-McVittie}, namely $R_+$, is given by \citep{far15}
\begin{equation}
R_{+} = \frac{1}{H(t)}\cos{\psi(t)} - \frac{1}{\sqrt{3}H(t)} \sin{\psi(t)}, 
\end{equation}
where $\sin{3 \psi(t)} = 3 \sqrt{3} m_0 H(t)$. For the Hubble factor given by Eq. \eqref{hubble-factor}, $R_+ \rightarrow \infty$ when $t \rightarrow \pm \infty$, and thus $R_+$ becomes a FLRW null infinity \citep{kal+10}.
  

Figure \ref{fig2} can be divided in three different regions:
\begin{itemize}
 \item Regular regions ($\theta_{\mathrm{in}} \theta_{\mathrm{out}} < 0$), painted in white.
 \item Trapped region ($\theta_{\mathrm{in}} \theta_{\mathrm{out}} > 0$, being $\theta_{\mathrm{in}} < 0$ and $\theta_{\mathrm{out}} < 0$), painted in light red.
 \item Anti-trapped region ($\theta_{\mathrm{in}} \theta_{\mathrm{out}} > 0$, being $\theta_{\mathrm{in}} > 0$ and $\theta_{\mathrm{out}} > 0$), painted in light blue.
\end{itemize}
 
\begin{figure}[htb]
\centering
\includegraphics[width=80mm,height=80mm]{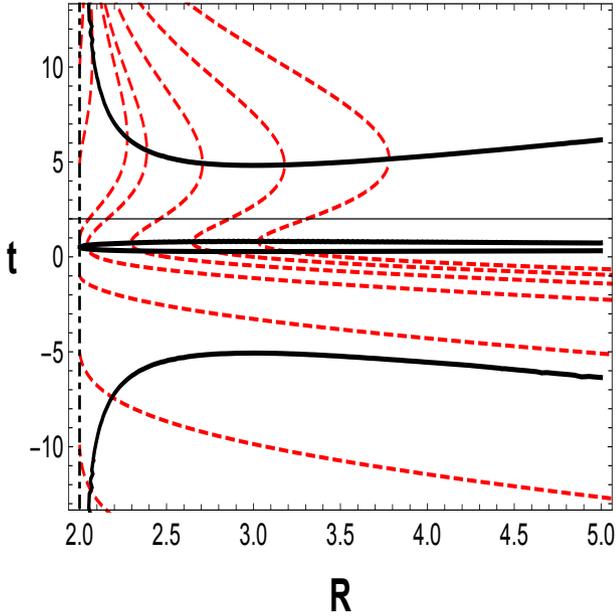} 
\caption{Radial ingoing null geodesics for McVittie metric in a bouncing cosmological model. The black lines denotes the trapping horizons and the dot-dashed line indicates the singularity. Here $p =1$, $m_0 = a_{b} = \sigma = 1$ and $w = 1/3$.} \label{fig3}
\end{figure}

\begin{figure}[htb]
\centering
\includegraphics[width=80mm,height=80mm]{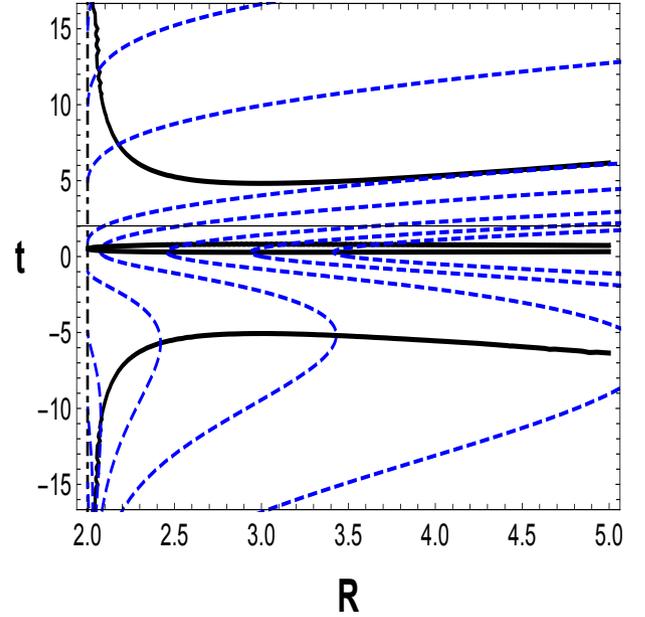} 
\caption{Radial outgoing null geodesics for McVittie metric in a bouncing cosmological model. The black lines denotes the trapping horizons and the dot-dashed line indicates the singularity. Here $p =1$, $m_0 = a_{b} = \sigma = 1$ and $w = 1/3$.} \label{fig4}
\end{figure}

We also compute the trajectories of ingoing and outgoing radial null geodesics by integrating the equation:
\begin{equation}\label{out-in}
 \frac{dR}{dt} = \sqrt{1- 2m_0/R} \left(H R \pm \sqrt{1- 2m_0/R}\right),
\end{equation}
 where the ``$-$'' (``$+$'') corresponds to the ingoing (outgoing) case. The results, shown in Figures \ref{fig3} and \ref{fig4}, are described briefly below:
 \begin{itemize}
 \item Ingoing geodesics expand to the past of the bounce . After the bounce ($t > 0$), they expand in the anti-trapped region until they cross the trapping horizon; once in the regular region of the spacetime $(dR/dt)_{\mathrm{in}} < 0$, they all seem to tend asymptotically to the surface  $R_{-} = 2m_{0}$, $t = \infty$, as shown in Figure \ref{fig3}. 
 \item Outgoing geodesics converge in the trapped region. After the bounce, outgoing null geodesics are always expanding ($(dR/dt)_{\mathrm{out}} > 0$) as depicted in Figure \ref{fig4}.
 \end{itemize}

\subsubsection{Case $p = 3$}

\begin{figure}[htb]
\centering
\includegraphics[width=70mm,height=70mm]{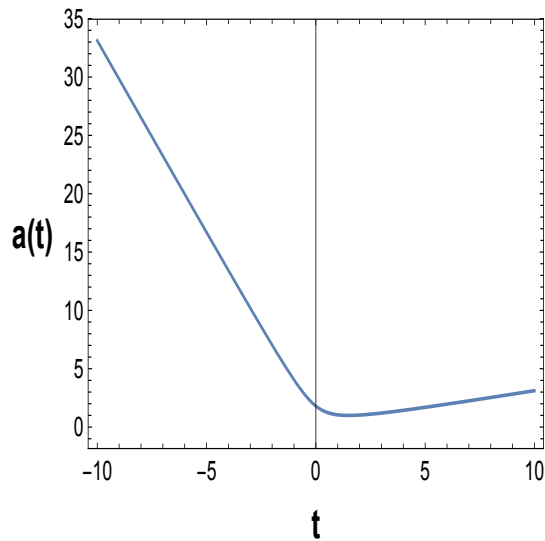} 
\caption{Plot of the scale factor as a function of time for $p = 3$. Here, $m_0 = a_{b} = \sigma = 1$ and $w = 1/3$.} \label{fig5}
\end{figure}

We now analyze the causal structure of the McVittie spacetime for $p = 3$. As can be seen in Figure \ref{fig5}, the slope of the scale factor after the bounce is much lower than for  $p =1 $ (see Fig. \ref{fig1}). The bounce occurs at larger positive values, more specifically at $t_{b} = 1.5$.

\begin{figure}[htb]
\centering
\includegraphics[width=80mm,height=80mm]{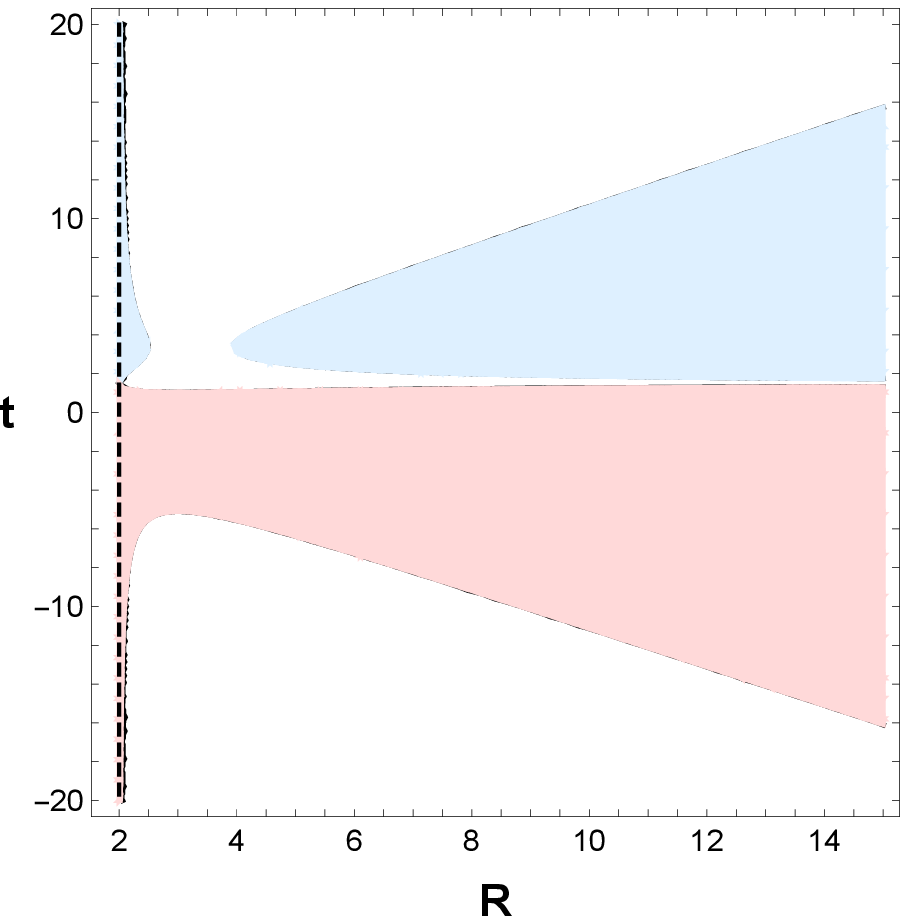} 
\caption{The black lines denotes the trapping horizons. The dot-dashed curve corresponds to the singularity. The white, red light and blue light  zones indicate the regular, trapped, and anti-trapped regions of the spacetime, respectively. Here $p =3$, $m_0 = a_{b} = \sigma = 1$ and $w = 1/3$.} \label{fig6}
\end{figure}

Inspection of Figure \ref{fig6} reveals that the effects of the asymmetry are more pronounce in the shape of the trapping horizons in the expanding phase. The most remarkable feature is that for $t > t_{b}$ the singularity is always cover by the inner trapping horizon. This is not the case for $p = 1$ as shown in Figure \ref{fig2}. We also note that in the contracting epoch the shape of the trapping horizons is almost identical for $p =1$ and $p =3$. A trapped region is present in the contracting phase, two separated anti-trapped regions occur in the expanding phase, and regular regions cover most of the spacetime for both large positive and negative values of the time coordinate.

\begin{figure}[htb]
\centering
\includegraphics[width=80mm,height=80mm]{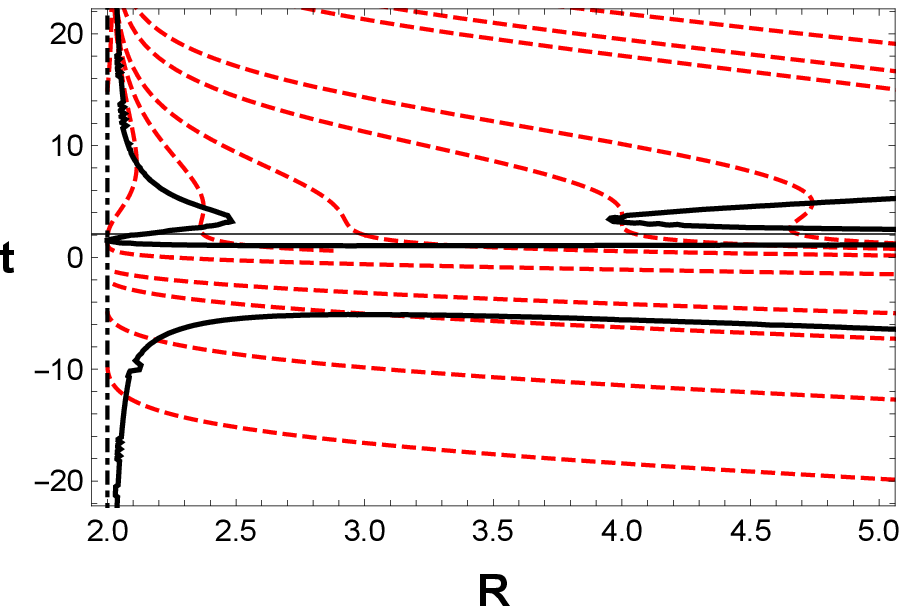} 
\caption{Radial ingoing null geodesics for McVittie metric in asymmetric bouncing cosmological model. The black lines denotes the trapping horizons and the dot-dashed line indicates the singularity. Here $p =3$, $m_0 = a_{b} = \sigma = 1$ and $w = 1/3$.} \label{fig7}
\end{figure}

\begin{figure}[htb]
\centering
\includegraphics[width=80mm,height=80mm]{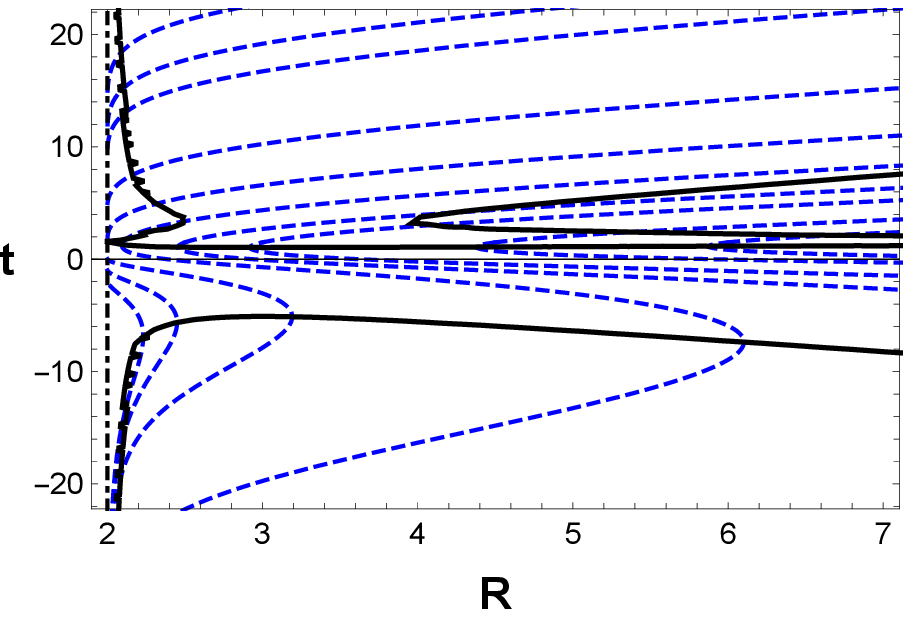} 
\caption{Radial outgoing null geodesics for McVittie metric in a bouncing cosmological model. The black lines denotes the trapping horizons and the dot-dashed line indicates the singularity. Here $p =3$, $m_0 = a_{b} = \sigma = 1$ and $w = 1/3$.} \label{fig8}
\end{figure}

The results of the integration of Equation \eqref{out-in} for ingoing and outgoing radial null geodesics are depicted in Figures \ref{fig7} and \ref{fig8}, respectively.

Ingoing geodesics have smaller and smaller radial coordinate as the bounce is approached. Some of these geodesics end up in the singularity while some others go through the bounce and enter in the expanding phase. We can see in Figure \ref{fig7} that all the ingoing geodesics in the regular region (white zone) seem to tend to the surface $R = 2 m_{0}$ as $t \rightarrow \infty$. 

Before the bounce, outgoing geodesics are expanding in the regular region. When they cross the trapping horizon (given by the condition $\theta_{\mathrm{out}} = 0$), they all converge in the trapped zone. While some geodesics have in their local future the singularity, some others are able to cross the bounce and expand either in the regular or in the anti-trapped region of the spacetime for $t > 0$.

\subsection{Trapping region before the bounce}\label{before-bounce}

We have shown that for both $p = 1$ and $p =3$, there exists a trapped region before the bounce that is bounded from below by  inner $(R_{-})$ and outer $(R_{+})$ trapping horizons for which $\theta_{\mathrm{out}} =0$. As the bounce is approached, both horizons get closer and at $t = t_{\mathrm{bf}} = -5.06$ for $p =1$ and $t_{\mathrm{bf}} = -5.09$ for $p =3$, they merge: only a trapping horizon exist for $R_{-} = R_{+} = 3 m_{0}$. For $t >  t_{\mathrm{bf}}$, the singularity becomes naked. 

The existence of a trapped region indicates that both ingoing and outgoing null geodesics are converging in that zone. Null rays that cross $R_{-}$ entering the trapping region are unable to turn around and escape This horizon, thus, acts as a one way membrane, hiding the singularity at $R = 2 m_{0}$. This analysis leads us to conclude that in the time interval $- \infty < t < t_{\mathrm{bf}}$ the solution contains a black hole.

\section{Discussion and conclusions}

In this work we analyze the causal structure of McVittie spacetime for an asymmetric bouncing cosmological background. We compute the location of the trapping horizons, and the trajectories of null radial ingoing and outgoing null geodesics through cosmic time for $p = 1$ and $p =3$. We find that a black hole is present since the beginning of the contracting phase in both cases. The inner trapping horizon $(R_{-})$ increases its radius as the contraction gathers pace. The range of values for the radial coordinate of the inner horizon $R_{-}$ is $2 m_{0}< R_{-} \le 3 m_{0}$. Ingoing and outgoing null geodesics that cross the surface $R_{-}$, enter the trapped zone of the spacetime. Close to the bounce, at $t = t_{\mathrm{bf}}$, inner  and outer trapping horizons merge, the black hole ceases to exist, and the solution exhibits a naked spacelike singularity. Trapping horizons appear again right before the bounce, and vanish right after it if $p=1$. Once the universe begins to expand inner $(R_{-})$ and outer $(R_{+})$ trapping horizons  again appear. This is not the case when the asymmetry in the scale factor is stronger, as for $p =3$.  Right before the bounce, trapping horizons appear again and persist forever on.

It remains to establish whether a black hole is present in the expanding phase. As previously mentioned, all ingoing geodesics in the regular region of the spacetime seem to tend to the surface $R_{-} = 2 m_0$, $t \rightarrow \infty$. Thus, if we are able to prove that i) null ingoing geodesics in the regular region cross the surface $R_{-} = 2 m_0$, $t \rightarrow \infty$ at a finite value of the affine parameter,  ii) such surface is regular, i.e. all the squared curvature invariants constructed with the Riemann tensor and its contractions are finite on it, and iii) once the geodesics transverse the surface  $R_{-} = 2 m_0$, $t = t_{\infty}$, they are in a trapped region, then the surface $R_{-} = 2 m_0$, $t \rightarrow \infty$ behaves as an event horizon and there is also a black hole in the expanding phase.\footnote{Kaloper and collaborators \citep{kal+10} were the first to show that the analysis of the behavior of ingoing null geodesics in the limit $t \rightarrow \infty$ reveals the presence of an event horizon in the McVittie spacetime.}

We point out that unlike all other McVittie models analyzed in the literature, there is no cosmological big singularity in the metric investigated in this work. In fact, the solution admits trajectories that never encounter a singularity, that is, they are geodesically complete. This peculiar feature of the model is related to the occurrence of the bounce.

The current solution can be improved by taking into account accretion of cosmological fluid by the central source. A possible approach to model such problem is considering the Generalized McVittie metric \citep{far15}. The corresponding energy-momentum is that of an imperfect fluid that naturally incorporates accretion. This is an issue that we shall explore in a future work.


\bibliography{apssamp.bib}%

\end{document}